\documentclass[%
preprint,
longbibliography,
 amsmath,amssymb,
aps,
]{revtex4-1}

\usepackage{graphicx}% Include figure files
\usepackage{dcolumn}% Align table columns on decimal point
\usepackage{bm}% bold math
\usepackage{subfig}
\usepackage{array}
\usepackage{tikz}
\usepackage[english]{babel}
\tikzstyle{long dashdotted}=[dash pattern=on 6pt off 3pt on \the\pgflinewidth off 3pt]

\graphicspath{{figures/}}
\usepackage{hyperref}% add hypertext capabilities
\begin{document}
\preprint{APS/123-QED}

\title{Mobility of bidisperse mixtures during bedload transport}% Force line breaks with \\

\author{R\'emi Chassagne}
 \email{remi.chassagne@inrae.fr}
 
\affiliation{%
 Univ. Grenoble Alpes, INRAE, IRSTEA, UR ETNA, 38000 Grenoble, France
}%
\author{Rapha\"el Maurin}%
\affiliation{%
 IMFT, Univ. Toulouse, CNRS - Toulouse, France
}%

\author{Julien Chauchat}
\affiliation{%
 Univ. Grenoble Alpes, LEGI, CNRS UMR 5519 - Grenoble, France
}%

\author{Philippe Frey}
\affiliation{%
 Univ. Grenoble Alpes, INRAE, IRSTEA, UR ETNA, 38000 Grenoble, France
}%

\date{\today}% It is always \today, today,

\begin{abstract}
The flow of segregated bidisperse assemblies of particles is of major importance for geophysical flows and bedload transport in particular. In the present paper, the mobility of bidisperse segregated particle beds was studied with a coupled fluid discrete element method. Large particles were initially placed above small ones and it was observed that, for the same flow conditions, the bedload transport rate is higher in the bidisperse configuration than in the monodisperse one. Depending on the Shields number and on the depth of the interface between small and large particles, different transport phenomenologies are observed, ranging from no influence of the small particles to small particles reaching the bed surface due to diffusive remixing. In cases where the small particles hardly mix with the overlying large particles and for the range of studied size ratios ($r<4$), it is shown that the increased mobility is not a bottom roughness effect, that would be due to the reduction of roughness of the underlying small particles, but a granular flow effect. This effect is analyzed within the framework of the $\mu(I)$ rheology and it is demonstrated that the buried small particles are more mobile than larger particles and play the role of a ``conveyor belt'' for the large particles at the surface. Based on rheological arguments, a simple predictive model is proposed for the additional transport in the bidisperse case. It reproduces quantitatively the DEM results for a large range of Shields numbers and for size ratios smaller than 4. The results of the model are used to identify four different transport regimes of bidisperse mixtures, depending on the mechanism responsible for the mobility of the small particles. A phenomenological map is proposed for bidisperse bedload transport and, more generally, for any granular flow on an erodible bed.
\end{abstract}


\maketitle

%\tableofcontents

\section{Introduction}
In mountain rivers, the sediment bed is generally composed of a large range of grain sizes. This polydispersity leads to size segregation, which is largely responsible for our limited ability to predict sediment flux \citep{bathurst2007, frey2011, dudill2018}. When segregating, small particles infiltrate the bed by kinetic sieving, falling down in holes formed by the matrix \cite{middleton1970}, and large particles rise to the bed surface \cite{savage1988}, resulting in inversely graded beds \cite{gray2018} which can be observed both in flume experiments and in the field. In 1914, Gilbert \cite{gilbert1914} was one of the first to observe experimentally that the introduction of finer sediments leads to an increase of sediment mobility. This has then been extensively studied due to strong implications for sediment transport and fluvial morphology \cite{hill2017, dudill2017, dudill2018}. The mobility of granular assemblies is also a key question in the study of several geophysical flows such as debris flows, pyroclastic flows, snow  avalanches and dune behavior. This, together with industrial applications, has led the granular community to study the influence of the slope \cite{mangeney2010, farin2014, maurin2018}, basal friction \cite{chedeville2014, edwards2017}, total volume \cite{staron2009} and polydispersity \cite{phillips2006, linares-guerrero2007, iverson2010, lai2017} on particle mobility.

Size segregation is often identified as the main mechanism responsible for the increased mobility of a polydisperse bed. In laboratory experiments with natural materials, Bacchi \textit{et al.} \cite{bacchi2014} showed that, due to kinetic sieving, small particles smooth the bed roughness and make the above large particles more mobile. In bedload transport laboratory experiments with a bidisperse bed, Dudill \textit{et al.} \cite{dudill2018} observed that the finer particles, after having infiltrated the first layers, drastically increased the sediment mobility. With two dimensional discrete element method simulations (DEM), Linares-Guerrero \textit{et al.} \cite{linares-guerrero2007} measured the run-out distance of dry bidisperse granular avalanches. They observed an increased mobility of the avalanche due to the presence of small particles segregating at the base of the granular flow and acting as a lubrication layer. Similarly, Lai \textit{et al.} \cite{lai2017} with DEM and laboratory experiments of granular collapse with fractal size distributions, observed the formation of a basal small particle layer increasing the total mobility. It seems therefore that size segregation, and in particular the formation of a small particle layer below large ones, plays an important role in the increased mobility process. Despite the few studies presented above, there is still no clear understanding of the physical mechanisms responsible for the increased mobility.

 Classically in bedload transport, bed mobility is interpreted in term of transport rate. The dimensionless transport rate, or Einstein parameter, defined as 
 \begin{equation}
 Q_s^* = \dfrac{Q_s}{((\rho^p/\rho^f-1)gd^3)^{1/2}},
 \label{eq:einstein}
 \end{equation}
 is related to the dimensionless fluid bed shear stress, or Shields number, defined as 
 \begin{equation}
 \theta = \dfrac{\tau_b^f}{(\rho^p-\rho^f)gd},
 \label{eq:shields}
 \end{equation}
 where $Q_s$ is the transport rate per unit width, $\rho^p$ (resp. $\rho^f$) is the particle density (resp. fluid density), $g$ is the gravity constant, $d$ is the bed surface particle diameter and $\tau_b^f$ is the fluid bed shear stress. Considering their physical meaning and the link with the transported granular layer, the representative diameter for both the Shields and the Einstein numbers should be taken as the surface layer particle diameter. It is classically chosen as the median surface diameter $d_{50}$ or $d_{84}$ ($84\%$ of the sediment is smaller than $d_{84}$) \cite{recking2013}. However, literature review \citep{bacchi2014, dudill2018, linares-guerrero2007, lai2017} underlines the importance of the depth structure in the mobility of the granular bed and, in particular, the influence of buried small particles. Therefore, understanding the impact of the bed depth structure on transport laws is of particular importance for an accurate description and prediction of turbulent bedload transport.
 
 While bedload transport has been mainly studied from the perspective of hydrodynamics, the present analysis illustrates the necessity to consider bedload as a granular phenomenon \cite{frey2011} and to describe the depth behaviour of the granular bed. In this paper, the mobility problem is therefore investigated from a granular perspective in the framework of the $\mu(I)$ rheology \cite{gdrmidi2004, jop2006, forterre2008}. For dense granular flows, the dry inertial number $I$ is the only dimensionless parameter controlling the system, where
 \begin{equation}
  I = \dfrac{d\dot{\gamma}}{\sqrt{P^p/\rho^p}},
  \label{eq:Idef}
 \end{equation}
with $\dot{\gamma}$ the shear rate and $P^p$ the granular pressure. The shear to normal granular stress ratio $\mu^p$ therefore depends only on the inertial number as
\begin{equation}
 \mu^p(I) = \dfrac{\tau^p}{P^p} = \mu_1 + \dfrac{\mu_2 - \mu_1}{I_0/I +1},
 \label{eq:muI}
\end{equation}
where $\tau^p$ is the granular shear stress and $\mu_1$, $\mu_2$ and $I_0$ are empirical coefficients fitted on dry experimental data. This rheology has been derived in monodisperse configurations and extended to bidisperse configurations in two dimensions \citep{rognon2007} and three dimensions \citep{tripathi2011}. In a recent work, Maurin \textit{et al.} \cite{maurin2016} studied the rheology of dense granular flows during bedload transport using a coupled fluid-DEM model. Despite the presence of water, they showed that the dry inertial number is still the controlling parameter. They found the $\mu(I)$ rheology to be valid in bedload transport over a wider range of inertial numbers and proposed another set of parameters than the one proposed by GDR Midi \cite{gdrmidi2004} with $\mu_1=0.35$, $\mu_2=0.97$ and $I_0=0.69$.\\

In the present paper, the mobility of bidispersed already segregated beds is studied from a granular perspective, considering coupled fluid-DEM simulations of turbulent bedload transport. This allows us to explain the modified mobility of a granular bed as a function of the granular depth structure, and to predict the sediment transport rate for polydisperse bedload transport. 
 
The numerical model is presented in section~\ref{sec:model}. The bed mobility is explored in section~\ref{sec:results}. Results are analysed within the $\mu(I)$ rheology framework in section~\ref{sec:mu_I} and an explanation for the increased mobility is presented. Based on rheological arguments, a simple predictive model for the additional transport is derived and compared with DEM simulations in section~\ref{sec:pred_model}. Finally the results are discussed in section~\ref{sec:discussion}.

\section{Numerical model and setup} \label{sec:model}

Our numerical model is a three dimensional discrete element method (DEM) using the open source code YADE \citep{smilaueretal.2015} coupled with a one-dimensional (1-D) turbulent fluid model. It has been derived and validated with particle-scale experiments \citep{frey2014} in \cite{maurin2015} and extended to bi-disperse configurations in \cite{chassagne2020}. It is briefly presented here but the interested reader should refer to Maurin \textit{et al.} \cite{maurin2015} for more details on the model and its validation. The DEM is a Lagrangian method based on the resolution of contacts. The inter-particle forces are modelled by a spring-dashpot system \citep{schwager2007} of stiffness $k_n$ in parallel with a viscous damper coefficient $c_n$ (corresponding to a restitution coefficient of $e_n=0.5$) in the normal direction; and a spring of stiffness $k_s$ associated with a slider of friction coefficient $\mu_g=0.4$ in the tangential direction. The values of $k_n$ and $k_s$ are computed in order to stay within the rigid grain limit \citep{roux2002, maurin2015}. The particles are additionally submitted to gravity, fluid buoyancy and turbulent drag force \cite{maurin2015}. Considering a particle $p$, the buoyancy force is defined as
\begin{equation}
 \bm{f}_b^p = -\dfrac{\pi d^{p3}}{6}\nabla P_{\bm{x}^p}^f,
\end{equation}
and the drag force as
\begin{equation}
 \bm{f}_D^p = \dfrac{1}{2} \rho^f \dfrac{\pi d^{p2}}{4}C_D ||\bm{u}_{\bm{x}^p}^f-\bm{v}^p||\left(\bm{u}_{\bm{x}^p}^f-\bm{v}^p \right),
 \label{eq:drag}
\end{equation}
where $d^p$ denotes the diameter of particle $p$, $\bm{u}_{\bm{x}^p}^f$ is the mean fluid velocity at the position of particle 
$p$, $P_{\bm{x}^p}^f$ is the hydrostatic fluid pressure at the position of particle $p$ and $\bm{v}^p$ is the velocity of particle $p$. The drag coefficient takes into account hindrance effects \citep{richardson1954} as $C_D = (0.4+24.4/Re_p)(1-\phi)^{-3.1}$, with $\phi$ the packing fraction and $Re_p = ||\bm{u}_{\bm{x}^p}^f-\bm{v}^p||d^p/\nu^f$ the particle Reynolds number, $\nu^f$ being the kinematic viscosity.

At transport steady state, the total granular phase (of small and large particles) only has a streamwise component with no main transverse or vertical motion. In such a case, the 3-D volume averaged equation for the fluid velocity reduces to a 1-D vertical equation in which the fluid velocity is only a function of the wall-normal component, $z$, and is aligned with the streamwise direction (see \cite{revil-baudard2013}) as
\begin{equation}
 \rho_f(1-\phi)\dfrac{\partial u_x^f}{\partial t} = \dfrac{\partial S_{xz}}{\partial z} + \dfrac{\partial R_{xz}}{\partial z} + \rho_f(1-\phi)g_x -n\left<f_{f_x}^p\right>^s,
\end{equation}
where $\rho_f$ is the density of the fluid, $S_{xz}$ is the effective fluid viscous shear stress of a Newtonian fluid of viscosity $\nu_f$. $R_{xz}$ is the turbulent fluid shear stress based on an eddy viscosity concept
\begin{equation}
 R_{xz} = \rho_f(1-\phi)\nu_t\dfrac{\partial u_x^f}{\partial z}.
 \label{eq:tSS}
\end{equation}
The turbulent viscosity  $\nu_t$ follows a mixing length approach that depends on the integral of the solid concentration profile to account for the presence of particles \citep{li1995}
\begin{equation}
\begin{array}{lr}
 \nu_t = l_m^2|\dfrac{\partial u_x^f}{\partial z}|, & l_m(z) = \kappa \displaystyle{\int_0^z}\dfrac{\phi_{max}-\phi(\zeta)}{\phi_{max}}d\zeta,
 \end{array}
\end{equation}
with $\kappa=0.41$ the Von-Karman constant and $\phi_{max} = 0.61$ the maximum packing of the granular medium (random close packing). The term $n\left<f_{f_x}^p\right>^s$ represents the momentum transfer associated with the interaction forces between fluid and particles. It is computed as the horizontal solid-phase average of the momentum transmitted by the drag force to each particle. 

The fluid model is classical in sediment transport \citep{drake2001, hsu2004, duran2012, revil-baudard2013, maurin2015, chauchat2018} and is only closed using a mixing length model and a closure for the drag force formulation. The latter are usual in the literature, and it has been shown in \citep{maurin2015a, maurin2015} that the results obtained in terms of granular behavior are very weakly sensitive to the fluid closure adopted.\\

\begin{figure}[h!]
    \centering
 \includegraphics[width=0.75\linewidth]{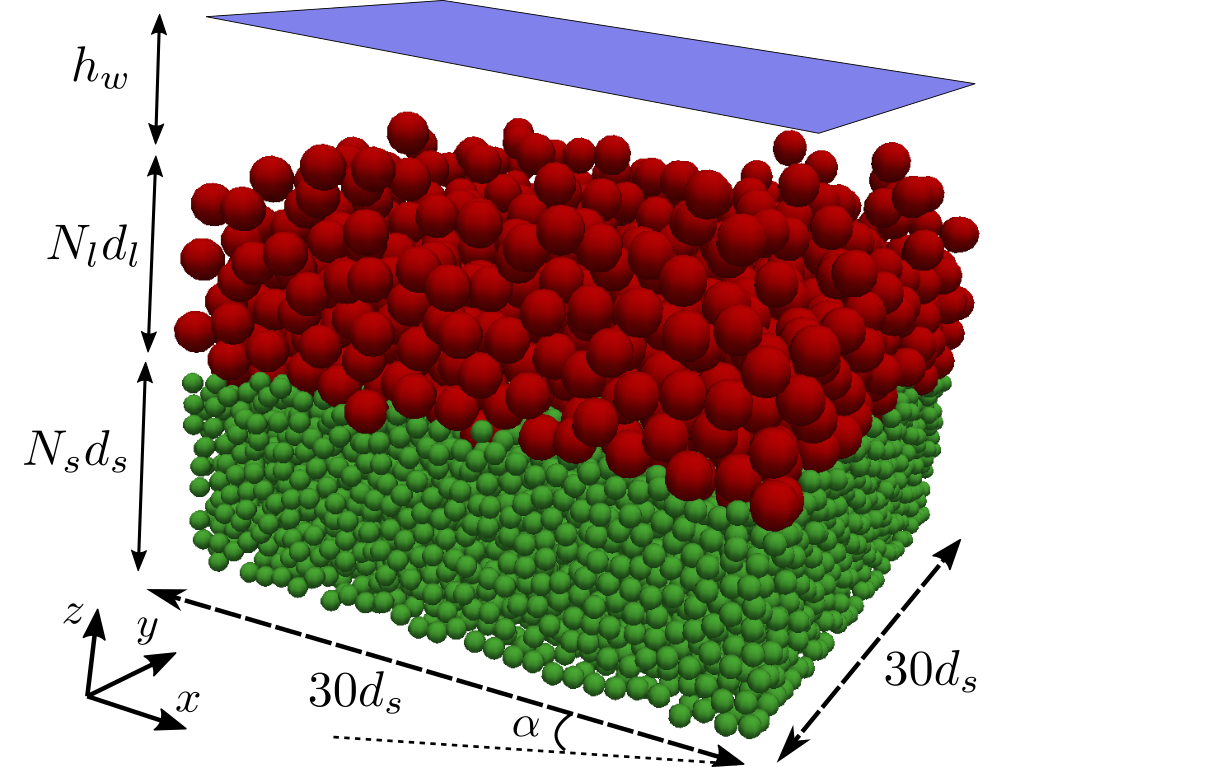}
 \caption{A typical numerical setup. Initially $N_l$ layers of large particles ($d_l=6$ mm) are deposited by gravity on $N_s$ layers of small particles ($d_s=3$ mm). The fluid of depth $h_w$ flows by gravity due to the slope angle $\alpha$ and entrains particles.}
 \label{fig:setup}
\end{figure}

\begin{figure}
 \centering
 \subfloat[]{\includegraphics[width=.3\linewidth]{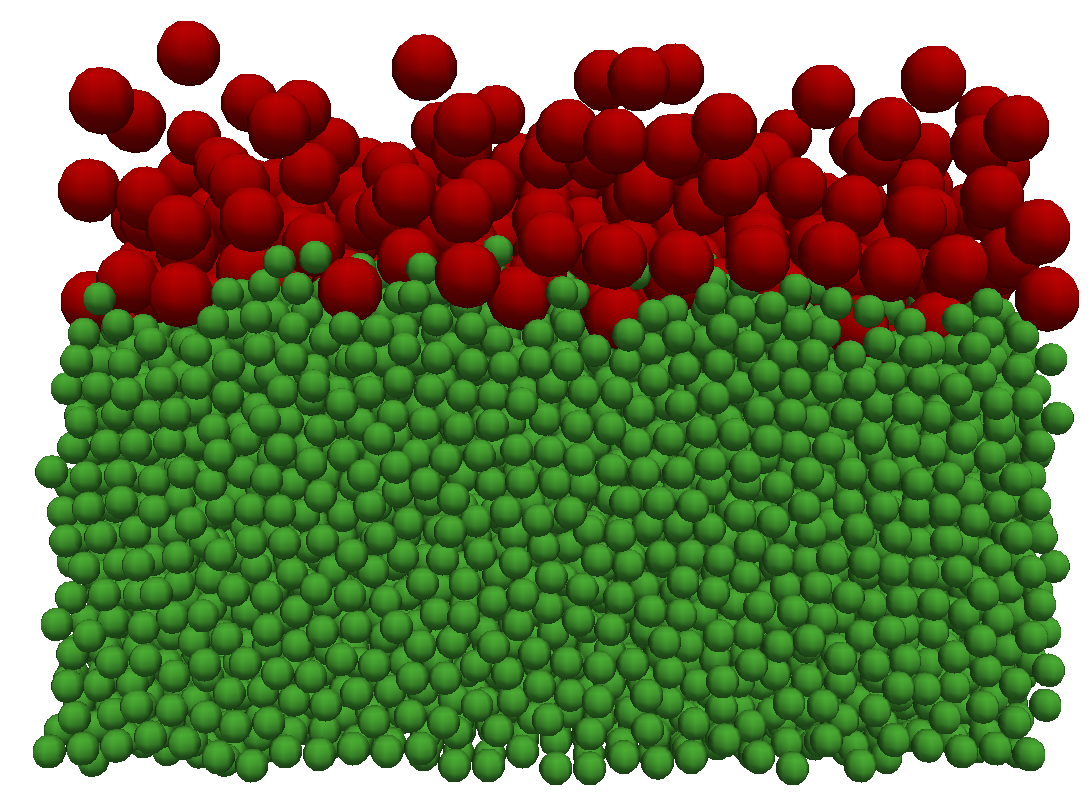}}
 \hspace{0.5cm}
 \subfloat[]{\includegraphics[width=.3\linewidth]{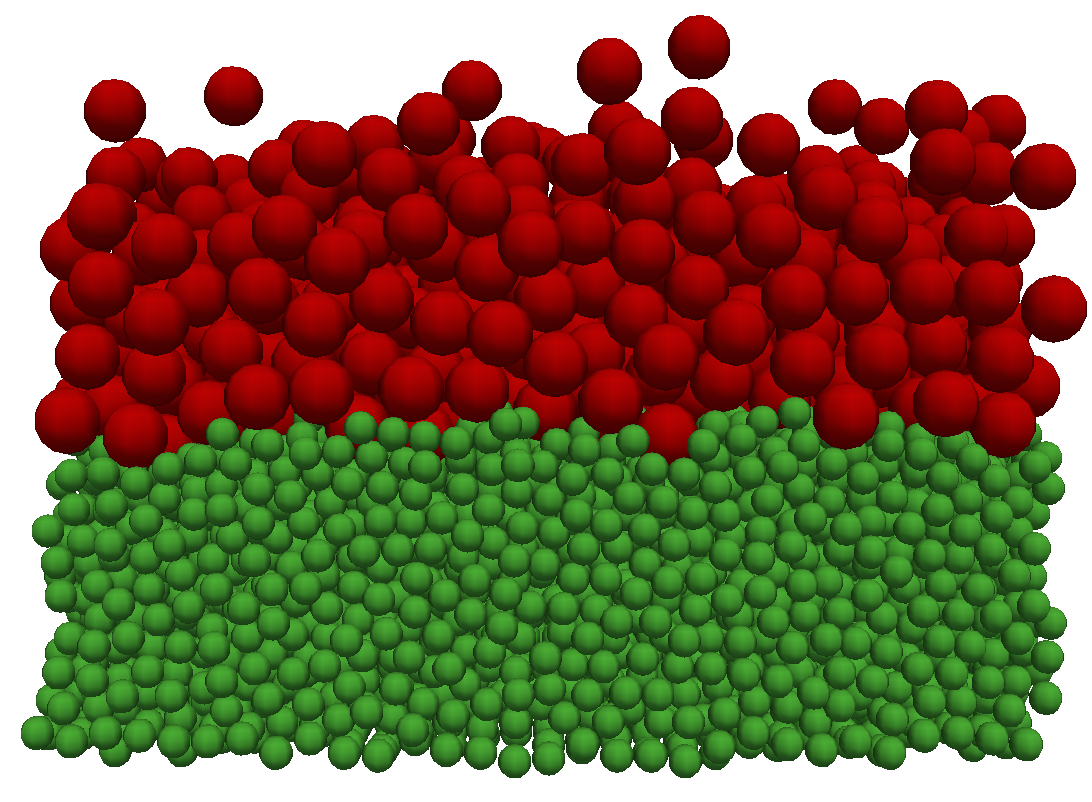}}
 \hspace{0.5cm}
 \subfloat[]{\includegraphics[width=.3\linewidth]{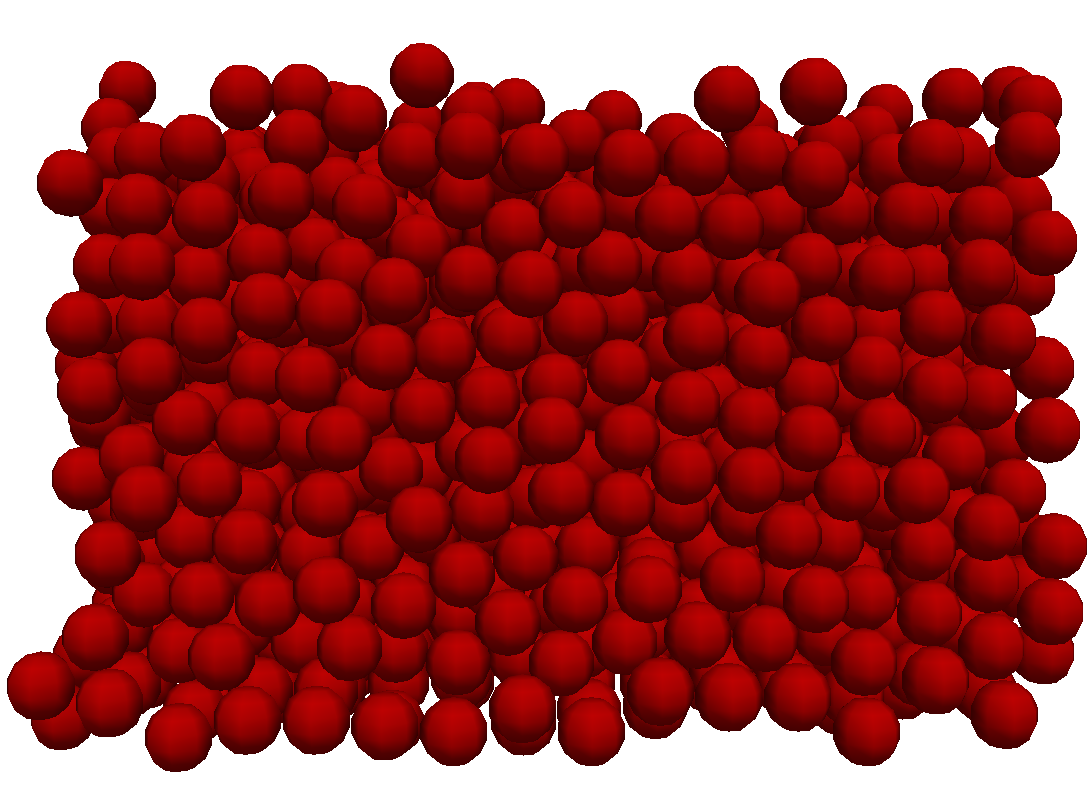}}
 \caption{Illustration of some considered configurations. (a) $N_l=2$, (b) $N_l=4$, (c) monidisperse case}
 \label{fig:config}
\end{figure}

The numerical setup is presented on figure~\ref{fig:setup}. In the following, subscripts $l$ and $s$ denote quantities for large and small particles respectively. Initially, small particles of diameter $d_s = 3$ mm and large particles of diameter $d_l=6$ mm are deposited by gravity over a rough fixed bed made of small particles. The size of the 3-D domain is $30d_s \times 30d_s$ in the horizontal plane in order to have converged average values \citep{maurin2015} and is periodic in the streamwise and spanwise direction. The number of particles of each class is assimilated to a number of layers, $N_s$ and $N_l$. They represent in terms of particle diameter the height that would be occupied by the particles  if the packing fraction was exactly $\phi_{max} = 0.61$, the maximal packing fraction. Equivalently, at rest, the volume occupied by large particles (resp. small particles) is $0.61\times30d_l\times30d_l\times N_ld_l$ (resp. $0.61\times30d_l\times30d_l\times N_sd_s$). Therefore, specifying $N_l$ and $N_s$ gives the number of particles in each class. The height of the bed at rest is thus defined by $H = N_sd_s + N_ld_l$.  The bed slope is fixed to $10\%$ ($\alpha=5.7^\circ$), representative of mountain streams. Since this study mainly focuses on cases where the bed surface is composed of only large particles, the Shields number definition is based on the large particle diameter as $\theta = \tau_f/((\rho^p-\rho^f)gd_l)$, where $\tau_f = \rho^f g h_w sin(\alpha)$ is the fluid bed shear stress, with $h_w$ the water depth. Simulations were performed for Shields numbers ranging from $0.1$ to $1$, i.e. from a few isolated particles transported at the bed surface to a ten grain thick mobile layer. Note that turbulent suspension never occured in our simulations. For each value of the Shields number, several configurations were considered with a varying number of layers of large particles $N_l = 1$, $2$, $3$ and $4$ which will be compared with a monodisperse large particle configuration considered as a reference case (see figure~\ref{fig:config}). In each case $N_s$ varied in order to keep the bed height $H$ constant equal to $H=8.5d_l$ for $\theta \leq 0.5$, $H=10.5d_l$ for $0.5 < \theta \leq 0.7$ and $H=16.5d_l$ for larger Shields numbers. This increase in the bed thickness was necessary in order to ensure an erodible bed bottom boundary condition. The origin of the vertical axis is set at the top of the particle bed at rest. The interface position, describing the transition between large and small particles, is therefore defined geometrically as $z_i = -N_ld_l$.

At the beginning of each simulation, the fluid flows by gravity and sets particles into motion. After approximately 20 seconds, a dynamical equilibrium is achieved between the fluid flow and the transport of sediment. The results are then time-averaged over a $280$s time period to ensure converged results. A mixed layer forms at the interface between small and large particles resulting from an equilibrium between diffusion and size segregation. The present study focuses on the relation between the fluid forcing and sediment transport once the steady state is achieved. Similarly to the Shields number, the Einstein parameter is defined with the large particle diameter as $Q_s^* = Q_s/\left((\rho^p/\rho^f-1)gd_l^3\right)^{0.5}$,
where $Q_s = \int_z \phi v_x^p dz$ is the transport rate per unit width, and $v_x^p$ is the bulk streamwise particle velocity. The horizontal averaged concentration of small (resp. large) particles is defined as $\phi_s$ (resp. $\phi_l$). By definition, the two concentrations sum to $\phi$ the total granular concentration,
\begin{equation}
 \phi_s + \phi_l = \phi.
\end{equation}

\section{Enhanced mobility due to bidispersity}\label{sec:results}

\begin{figure}[h!]
 \centering
 \subfloat[]{\includegraphics[height=6cm]{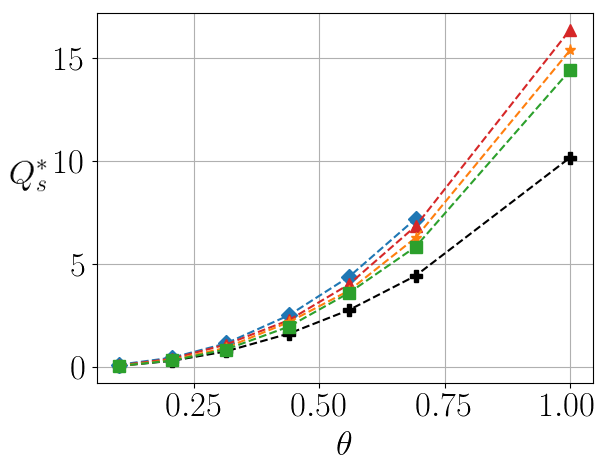}}
 \subfloat[]{\includegraphics[height=6cm]{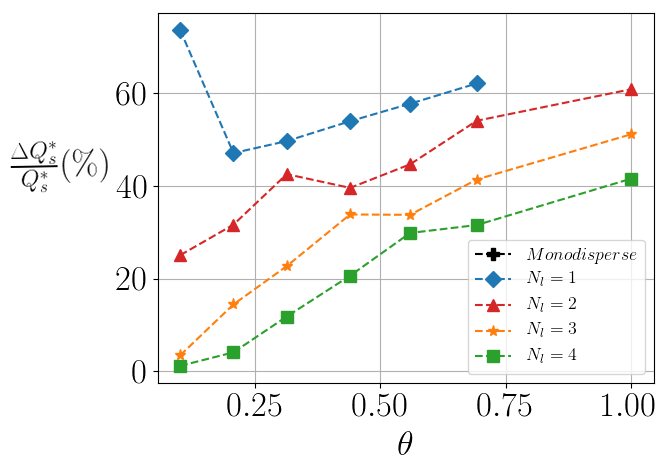}}
 \caption{(a) Solid transport rate as a function of the Shields number for all simulations. (b) Increased transport rate in percentage compared with monodisperse configuration.}
 \label{fig:transport_law}
\end{figure}

In figure~\ref{fig:transport_law}a is plotted the steady state dimensionless solid transport rate as a function of the Shields number. In all configurations, the dimensionless transport rate increases with the Shields number. The transport rate is remarkably stronger in all bidisperse configurations with respect to the monodisperse case, evidencing enhanced particle mobility. Figure~\ref{fig:transport_law}b shows the bidisperse transport relative to monodisperse configurations, increasing up to $50\%$. The increase of transport is almost linear with the Shields number and is stronger when the number of layers of large particles  $N_l$ is small. Indeed, for a lower $N_l$, small particles are closer to the surface (see figure~\ref{fig:config}) and are more likely to influence transport. This indicates that the depth of the interface between large and small particles, $z_i$, plays a role in the transport efficiency. At low Shields numbers and for $N_l=4$, almost no increase of transport is observed. In that case, the interface position is too deep to affect the bed mobility, and the bidisperse bed behaves as if it were monodisperse. 
%Note that, by definition, figure~\ref{fig:transport_law}b also represents the error that one would make by predicting the transport in the bidisperse case with a monodisperse transport law. 
Overall, without modification of the fluid forcing, a substantial increase of transport is observed just by changing the particle size in the bed depth profile. 

\begin{figure}[h]
\centering
 \subfloat[$\theta \sim 0.2$, $N_l=4$]{\includegraphics[height=4.5cm]{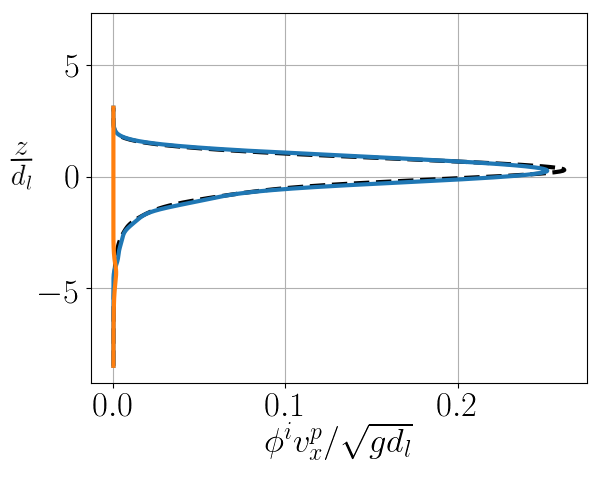}}
 %\subfloat[$\Theta \sim 0.45$, $N_l=4$]{\includegraphics[height=4.5cm]{qsN4sh03}}
 %\subfloat[$\Theta \sim 0.2$, $N_l=2$]{\includegraphics[width=0.45\linewidth]{qsN2sh01}}
 \subfloat[$\theta \sim 0.45$, $N_l=2$]{\includegraphics[height=4.5cm]{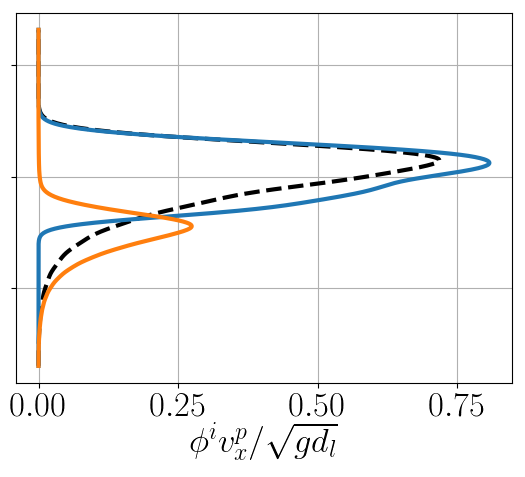}}
 \subfloat[$\theta \sim 0.55$, $N_l=1$]{\includegraphics[height=4.5cm]{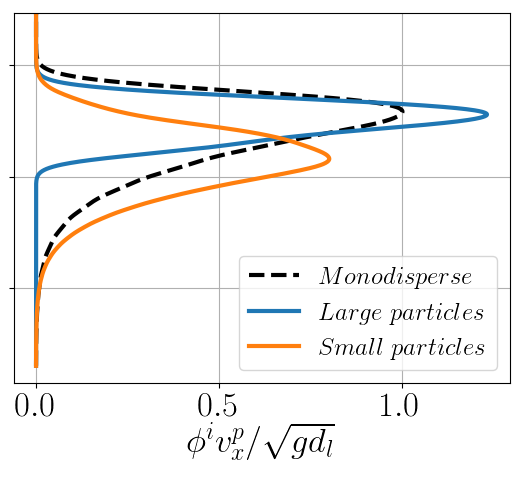}}
 \caption{Transport profiles of each class of particles for different configurations and Shields numbers. The transport profile of the large particles in the monodisperse case for the same Shields number is also plotted for comparison.  Note changes in the abscissa scale.}
 \label{fig:transport_profiles}
\end{figure}

To expand the transport description, the local transport rate of each class of particle is defined as $q_s^i(z) = \phi_i(z)v_x^p(z)$, where $\phi_i(z)$ is the concentration of particle class $i=l, s$. Figure~\ref{fig:transport_profiles} shows the local transport rate depth profile of each class of particles for different typical configurations. The transport rate of large particles in the monodisperse case is also plotted in black dashed line for comparison. For $\theta \sim 0.2$ and $N_l=4$ (figure~\ref{fig:transport_profiles}a), almost no increase of transport ($\sim 4\%$) is observed, and the small particles are barely transported. Increasing the Shields number, figure~\ref{fig:transport_profiles}b shows that the small particles are transported but remain buried in the bed. When comparing the transport rate profile of small particles with the monodisperse configuration (dashed line), the small particle transport is higher than the large one at the same depth. The same observation is true for the overlying large particles. The total transport, being the sum of both the small and large particle transport, is therefore much higher in the bidisperse case than in the monodisperse case.  For $\theta \sim 0.55$ and $N_l=1$ (figure~\ref{fig:transport_profiles}c) the transport of small particles is even stronger and small particles are present up to the bed surface, while they remained buried in the previous configuration (figure~\ref{fig:transport_profiles}b). It is therefore possible to draw two main conclusions. First, the observed increase of transport is a direct consequence of the mobility of the small particles. Second, even the large particle transport is significantly higher than in the monodisperse case.

\begin{figure}
 \centering
 \includegraphics[width=.5\linewidth]{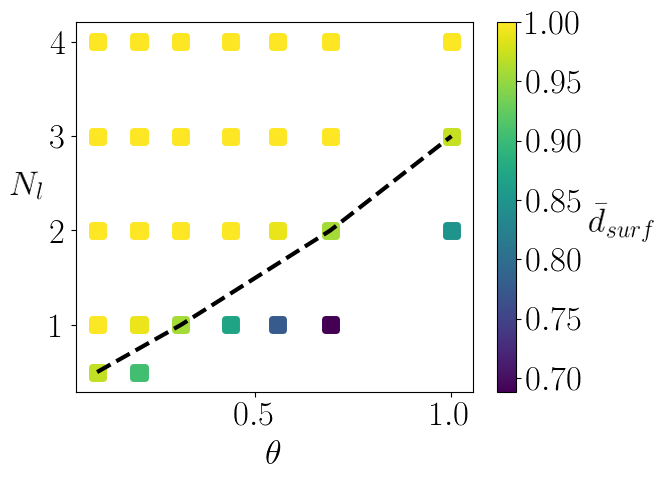}
 \caption{Mean surface diameter as a function of the Shields number and the large particle number of layers. The dashed line shows the transition between a large particle surface state to a mixture surface state.}
 \label{fig:dsurf}
\end{figure}

Two types of phenomenology are observed in the results. On the one hand small and large particles remain well separated, with small particles buried deep in the bed (figure~\ref{fig:transport_profiles}a, b). On the other hand, small and large particles are mixed at the surface (figure~\ref{fig:transport_profiles}c). The width of the transition between small and large particles depends on the relative importance of segregation over diffusion, the ratio of which can be defined as the Peclet number $P_e$ \citep{chassagne2020}. If diffusion is strong enough compared to segregation, small buried particles can reach the surface. To characterise the surface state, the surface diameter is computed as the mean particle diameter above $z=0$ as
\begin{equation}
 d_{surf} = \dfrac{\int_0^{+\infty}\phi_s(z)d_s + \phi_l(z)d_ldz}{\int_0^{+\infty}\phi_s(z)+\phi_l(z)dz}.
\end{equation}
The non-dimensional surface diameter is set between $0$ (only small particles at surface) and $1$ (only large particles) with the following transformation
\begin{equation}
 \bar{d}_{surf} = \dfrac{d_{surf}-d_s}{d_l-d_s}.
 \label{eq:dsurf}
\end{equation}
Figure~\ref{fig:dsurf} shows in scatter plot the value of the surface diameter as a function of the Shields number and the number of layers of large particles. The domain is clearly separated into two parts deliminated by the dashed line. Above the dashed line, the bed surface is only composed of large particles while below it is composed of a mixture of both small and large particles. For a given value $N_l$, there exists a transition Shields number $\theta_t(N_l)$ which separates a monodisperse from a bidisperse bed surface. For $\theta<\theta_t$, diffusion is weak compared to segregation, while for $\theta>\theta_t$ it is strong enough to move small particles up to the bed surface. This therefore indicates that the Peclet number $P_e$ depends on the Shields number. In addition $\theta_t$ increases with $N_l$. Indeed, when $N_l$ increases, the transition depth $z_i$ between small and large particles is deeper in the bed and diffusion needs to be even stronger for the small particles to reach the surface. 
For $N_l=4$ the surface is always composed of large particles. There is no doubt that increasing again the Shields number will eventually bring small particles at the surface. Two simulations for $N_l=0.5$ have also been plotted for illustration. By definition in these cases, the bed surface is necessarily composed of a mixture of large and small particles. 

In cases where the surface is composed of a mixture of small and large particles (below the dashed line), the increased transport can be attributed to a fluid effect. Indeed, at constant fluid shear stress, the ratio between the drag force contribution and the buoyant weight is inversely proportional to the diameter, so that the transport rate is \textit{a priori} higher for a mixture surface state. In cases where the small and large particles are well separated (above the dashed line), the increased transport rate cannot be attributed to a fluid effect. Indeed the length over which the fluid shear stress is fully transferred to the granular bed is much smaller than the grain size (see \cite{ouriemi2009}, \cite{maurin2018}), and it is verified in appendix~\ref{sec:app1} that it is indeed fully transferred to the granular bed below $z=0$. The increased transport is therefore necessarily due to a granular process. In the next section, the study focuses only on the configurations where small and large particles are well separated and where the bed surface is composed only of large particles. The granular process responsible for the increase of mobility is investigated through a mechanical analysis of the granular bed properties.  

\section{Interpretation as a granular process}\label{sec:mu_I}

% \begin{figure}[h]
%  \centering
%  \includegraphics[width=0.5\linewidth]{stressN2sh03}
%  \caption{Granular pressure (full line) and granular shear stress (dotted line) for $\Theta=0.45$ and for the $N_l=2$ and the monodisperse configuration.}
%  \label{fig:stress}
% \end{figure}
The granular stress tensor can be computed from the DEM. Considering a horizontal slice of volume $V$, the granular stress tensor is calculated as \citep{goldhirsch2010, andreotti2013}
\begin{equation}
 \sigma_{ij}^p = -\dfrac{1}{V}\sum_{p \in V}m^{p}v_i^{\prime p}v_j^{\prime p} 
- \dfrac{1}{V}\sum_{c \in V}f_i^{c}b_j^c,
\end{equation}
where the sum is performed over the ensemble of particles $p$ and contacts $c$ inside the volume $V$, $v_k^{\prime p}= v_k^{p}- \left<v_k^p\right>^s$ is the $k$ component of the spatial velocity fluctuation of particle $p$, $\bm{f}^c$ is the interaction force at contact $c$ on particle $\alpha$ by particle $\beta$ and $\bm{b}^c=\bm{x}^{\beta}-\bm{x}^{\alpha}$ is the branch vector. Due to the one dimensional structure of the flow, Maurin \textit{et al.} \cite{maurin2015a, maurin2016} showed that, in the steady state bedload configuration, $\sigma_{zz}^p = Tr(\sigma^p)/3$ and the only non diagonal term which is non null is $\sigma_{xz}^p$. The granular stress can therefore be described by only two scalar parameters which are the granular pressure $P^p = \sigma_{zz}^p$ and the shear stress $\tau^p=\sigma_{xz}^p$.\\

\begin{figure}
 \subfloat[]{\includegraphics[height=5.5cm]{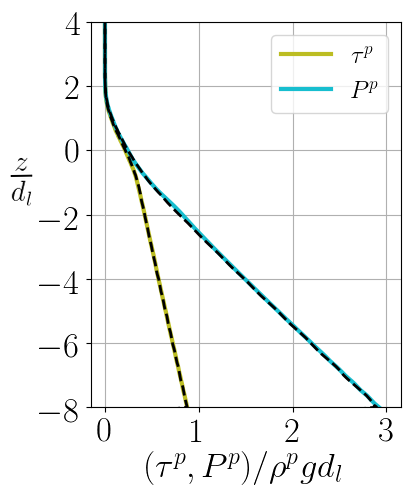}}
 \subfloat[]{\includegraphics[height=5.5cm]{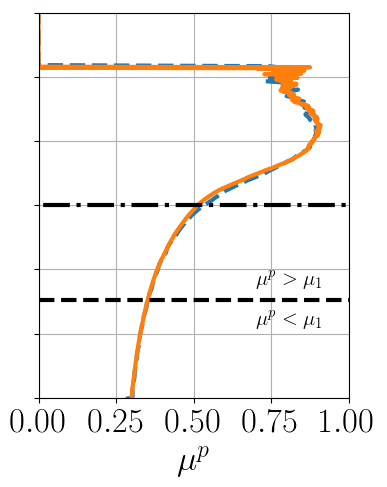}}
 \subfloat[]{\includegraphics[height=5.5cm]{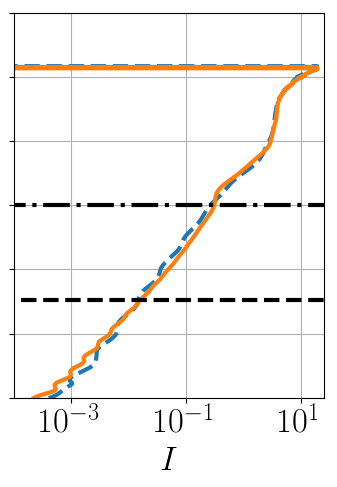}}
 \subfloat[]{\includegraphics[height=5.5cm]{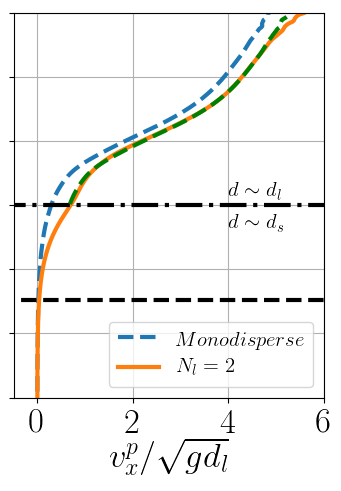}}
 \caption{Comparison of the monodisperse (dotted line) and the bidisperse $N_l=2$ (full line) configuration for $\theta \sim 0.45$. (a) Pressure and shear stress profiles, (b) friction coefficient profiles, (c) inertial number profiles and (d) velocity profiles. The dotted green line corresponds to a translation of $\Delta v = 0.383\sqrt{gd_l}$ of the monodisperse velocity profile (blue dotted line). The lower horizontal line at $z_1$  (\protect\tikz[baseline]{\protect\draw[line width=0.3mm,densely dashed] (0,.5ex)--++(0.65,0) ;)}) separates the quasi-static regime from the flowing dense regime. The upper horizontal line at $z_i$
 (\protect\tikz[baseline]{\protect\draw[line width=0.3mm,long dashdotted] (0,.5ex)--++(0.7,0) ;)}) shows the transition from small to large particles in the bidisperse configuration.}
 \label{fig:muvx}
\end{figure}

Figure~\ref{fig:muvx}a compares, for $\theta \sim 0.45$, the monodisperse and the bidisperse ($N_l=2$) components of the stress tensor. The pressure and the shear stress exhibit the same behavior in the monodisperse and bidisperse configurations. For the same forcing, the response of the bed in terms of granular stresses is therefore the same whatever the constitution of the bed. However, the transport profiles (figure~\ref{fig:transport_profiles}b) show that the bidisperse bed is more mobile than the monodisperse one. This means that the dynamical response is dependent on the bed composition. This is analysed within the framework of the $\mu(I)$ rheology, relating the friction coefficient $\mu^p = \tau^p/P^p$ to the inertial number $I$. The diameter to consider in the expression of the inertial number~\eqref{eq:Idef} is the local volume-averaged diameter \citep{rognon2007, tripathi2011} $d=\phi_sd_s+\phi_ld_l$ (which simplifies to $d=d_l$ in the monodisperse case). Following GDR Midi \cite{gdrmidi2004}, the rheology of dense granular flows can be seen as follows. If $\mu^p \leq \mu_1$, where $\mu_1$ is the static friction coefficient, no motion is observed and $I=0$. If $\mu^p>\mu_1$, there exists a one to one correspondence between the friction cofficient $\mu^p$ and the inertial number $I$. 

The friction coefficient is plotted in figure~\ref{fig:muvx}b and, as expected from the similarity of the granular stress profiles (figure~\ref{fig:muvx}a), it is the same in the bidisperse and the monodisperse configuration. As a consequence, the inertial number profiles should be the same in both configurations and that is indeed the case as observed in figure~\ref{fig:muvx}c. The dashed line (\protect\tikz[baseline]{\protect\draw[line width=0.3mm,densely dashed] (0,.5ex)--++(0.65,0) ;)}), defines a depth $z_1$ such that $\mu^p(z_1) = \mu_1$, the theoretical transition between static and dense granular flows. The dashed-dotted line (\protect\tikz[baseline]{\protect\draw[line width=0.3mm,long dashdotted] (0,.5ex)--++(0.7,0) ;)}) shows the interface depth $z_i$ between small and large particles.

Figure~\ref{fig:muvx}d shows the bulk particle velocity for both configurations. For $\mu<\mu_1$ or equivalently $z<z_1$, the inertial number and the velocity are indeed small but not exactly zero. This is due to non-local effects, that the $\mu(I)$ rheology is not able to capture \cite{kamrin2012, bouzid2013}. It corresponds to a quasi-static flow, or creeping regime, in which the velocity is exponentially decreasing into the bed (\cite{ferdowsi2017}, \cite{chassagne2020}). 
In order to understand the increased mobility in the bidisperse configuration, the quasi-static regime is assumed to have a negligible impact on transport and is not considered in this study. For $z>z_1$, as the friction coefficient is similar in both configurations (see figure~\ref{fig:muvx}b), the inertial number is also supposed to be the same
\begin{equation}
 I_{b} = I_{m},
 \label{eq:Iequal}
\end{equation}
where subscript $b$ (resp. $m$) denotes the bidisperse (resp. monodisperse) configuration. For $z_1<z<z_i$, the particle diameter in the bidisperse simulation is $d_b \sim d_s$, and $d_m = d_l$ for the monodispserse case. Equation~\ref{eq:Iequal} becomes
\begin{equation}
 \dfrac{d_s\dot{\gamma}_b}{\sqrt{P^p/\rho^p}} \sim \dfrac{d_l\dot{\gamma}_m}{\sqrt{P^p/\rho^p}}.
 \label{eq:Iequal2}
\end{equation}
The granular pressure being the same in both configurations (see figure~\ref{fig:muvx}a), gives
\begin{equation}
 \dot{\gamma}_b \sim \dfrac{d_l}{d_s}\dot{\gamma}_m.
 \label{eq:Iequal3}
\end{equation}
Integrating equation~\ref{eq:Iequal3} from $z_1$ to $z \leq z_i$, and assuming that the velocities are zero in $z_1$, yields
\begin{equation}
 v_b^p(z) \sim  \dfrac{d_l}{d_s} v_m^p(z),
\end{equation}
and therefore the velocity is higher in the bidisperse case than in the monodisperse case. This is perfectly observed in figure~\ref{fig:muvx}d. It means that for the same granular stress state, small particles are transported more easily than larger particles.

For $z>z_i$, the particle diameter is $d_l$ in both configurations and equation~\eqref{eq:Iequal2} simplifies to
\begin{equation}
 \dot{\gamma}_b \sim \dot{\gamma}_m,
\end{equation}
and by integration from depth $z_i$ to $z$,
\begin{equation}
 v_b^p(z) \sim v_m^p(z) + \left(v_b^p(z_i)-v_m^p(z_i)\right) \sim v_m^p(z) + \Delta v,
\end{equation}
meaning that the particle velocity profile in the bidisperse case is just a translation of the velocity profile in the monodisperse case. In figure~\ref{fig:muvx}d is plotted, in the upper part of the bed, $v_m^p(z)+\Delta v$, with $\Delta v = 0.383\sqrt{gd_l}$ measured in the DEM simulation. The obtained curve is completely superimposed on the velocity profile in the bidisperse configuration. In both configurations, the large particles at the top have exactly the same behaviour. \\

The proposed granular analysis explains the observation made previously in figure~\ref{fig:transport_profiles}, in which a layer of small particles was observed to be transported faster than larger particles at the same depth. Small particles consequentely play the role of a conveyor belt for the overlying particles and $\Delta v$ represents a slip velocity. It additionally shows that the enhanced mobility is not a roughness effect, due to the reduction of roughness by smaller particles below the large particle layer. Indeed, if particles do not move at the interface, $\Delta v = v_b^p(z_i)-v_m^p(z_i)$ is zero and no enhanced mobility is observed, as in figure~\ref{fig:transport_profiles}a. The fluid origin for the increased mobility can be discarded because the fluid shear stress is already fully transferred to the granular shear stress below $z=0$ (see appendix~\ref{sec:app1}). This analysis confirms that the enhanced mobility originates in the granular rheological properties of bidisperse beds.\\

This rheological analysis gives a qualitative understanding of the granular bed behaviour in the bidisperse configuration. To be more quantitative, the previous conclusions are used to predict analytically the additional transport in the bidisperse case.
% More generally, without assuming that the small and large particles are splitted into two distincts regions without mixing, the equality of the inertial numbers~\eqref{eq:Iequal} can be rewritten as
% \begin{equation}
%  d(z)\dot{\gamma}_b(z) = d_l\dot{\gamma}_m(z),
% \end{equation}
% with $d(z) =\phi_s(z)d_s+\phi_l(z)d_l$ and by integration
% \begin{equation}
%  v_b(z) = \int_0^z\dfrac{d_l}{d(z^{\prime})}\dot{\gamma}_m(z^{\prime})dz^{\prime}
% \end{equation}

\section{A predictive model for the additional transport}\label{sec:pred_model}

In this section, a simple model is derived, the purpose of which is to predict the additional transport observed in the bidisperse case. To obtain a predictive model, the additional transport will be expressed as a function of the monodisperse quantities ($\phi_m$, $v_m^p$, etc...). The configuration is ideally simplified as a two layer problem in which small and large particles are completely separated at the interface depth $z_i$. The mixed layer of small and large particles, observed in the bidisperse DEM simulations, is here neglected. Therefore it is assumed that the mixture concentration profiles are identical in the bidisperse and in the monodisperse configuration, ie. $\phi_m(z)=\phi_b(z)$.

% \begin{figure}
%  \includegraphics[width=.5\linewidth]{schema_mod}
%  \caption{}
%  \label{fig:setup_model}
% \end{figure}

The transport in the bidisperse case is expressed as,
\begin{equation}
Q_b = \int_{-\infty}^{+\infty} v_b^p(z)\phi_b(z)dz.
\label{eq:minimodel1}
\end{equation}
Below the interface between large and small particles, i.e. $z\leq z_i$, the previous analysis has shown that $v_b^p(z) = d_l/d_s v_m^p(z)$, while for $z>z_i$, $v_b^p(z) = v_m^p(z) + \Delta v$. Splitting the integral into two parts, below and above $z_i$, placing the velocity expression into equation~\ref{eq:minimodel1} and recalling that $\phi_b(z) = \phi_l(z)+\phi_s(z)$ is assumed to be equal to $\phi_m(z)$, one obtains
\begin{equation}
 Q_b = \int_{-\infty}^{z_i} \dfrac{d_l}{d_s}v_m^p(z)\phi_m(z)dz + \int_{z_i}^{+\infty} (v_m^p(z)+\Delta v)\phi_m(z)dz.
 \label{eq:minimodel3}
\end{equation}
Distributing the second term and combining it with the first term, it comes
\begin{equation}
Q_b = Q_m + (\dfrac{d_l}{d_s}-1)\int_{-\infty}^{z_i} v_m^p(z)\phi_m(z)dz + \int_{z_i}^{+\infty}\Delta v\phi_m(z)dz.
\label{eq:minimodel4}
\end{equation}
where $Q_m = \int_{-\infty}^{+\infty} v_m^p(z)\phi_m(z)dz$ is the monodisperse transport rate. Recalling that $\Delta v$ is independent of $z$, the additional transport due to the presence of small particles can therefore be expressed as
\begin{equation}
\Delta Q = (\dfrac{d_l}{d_s}-1)\int_{-\infty}^{z_i} v_m^p(z)\phi_m(z)dz + \Delta v\int_{z_i}^{+\infty}\phi_m(z)dz = \Delta Q_1 + \Delta Q_2.
\label{eq:minimodel6}
\end{equation}
The term $\Delta Q_1$ represents the additional transport below the interface of the small particles, more mobile than larger particles. The term $\Delta Q_2$ represents the additional transport of the large particles at the surface due to the conveyor belt effect. Note that in the monodisperse limit (i.e. $d_s = d_l$), both terms vanish. This is obvious for $\Delta Q_1$. For $\Delta Q_2$, it is $\Delta v = v_b^p - v_m^p$, which cancels in the monodisperse limit ($v_b^p = v_m^p$). Note that the additional transport in the bidisperse configuration (equation~\eqref{eq:minimodel6}) is expressed only as a function of monodisperse variables.\\

\begin{figure}
\centering
 \subfloat[]{\includegraphics[height=6cm]{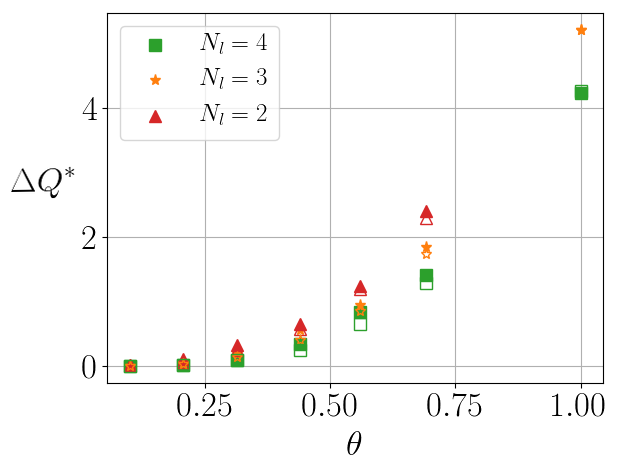}}
 \subfloat[]{\includegraphics[height=6cm]{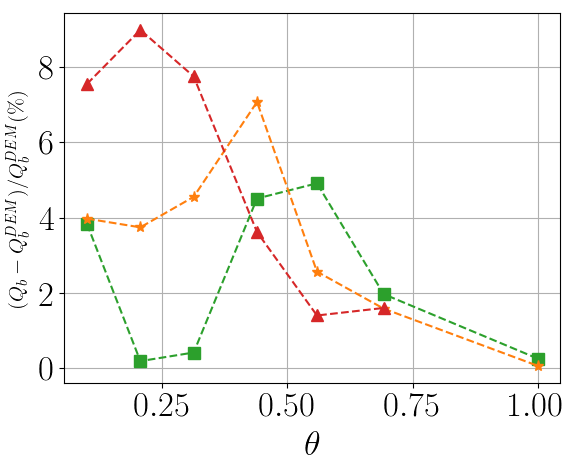}}
 \caption{(a) Dimensionless additional transport measured in the DEM simulations (full symbols) and predicted by equation~\eqref{eq:minimodel6} (empty symbols), for different values of the Shields number and $N_l$. Only cases for which the surface is exclusively composed of large particles are presented for readability. (b) Error between the total transport predicted by equation~\eqref{eq:minimodel6} and the transport computed with the DEM simulations.}
 \label{fig:res}
\end{figure}

In order to verify that the model is consistent with the transport mechanisms at play, equation~\eqref{eq:minimodel6} is first tested using DEM monodisperse simulations as inputs. The additional transport terms $\Delta Q_1$ and $\Delta Q_2$ are computed using the DEM velocity and concentration profiles $v_m^p$, $\phi_m$ and estimating the slip velocity $\Delta v$ directly on the DEM simulations. %as $\Delta v= 1/z_i\int_{z_i}^0(v_b^p(z) - v_m^p(z))dz$. 
The predicted dimensionless additional transport rates are plotted in figure~\ref{fig:res}. The additional transport in the bidisperse case is very well predicted by equation~\eqref{eq:minimodel6} for all values of Shields number and for all numbers of layers of large particles. The small errors obtained with equation~\eqref{eq:minimodel6} show that the model contains the significant physical ingredients acting in this transport process.\\

\begin{figure}
 \centering
 \subfloat[]{\includegraphics[height=5.5cm]{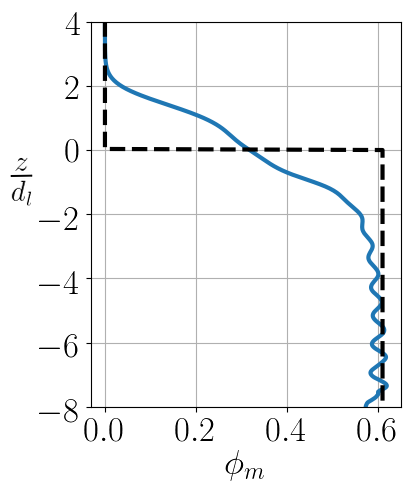}}
 \subfloat[]{\includegraphics[height=5.5cm]{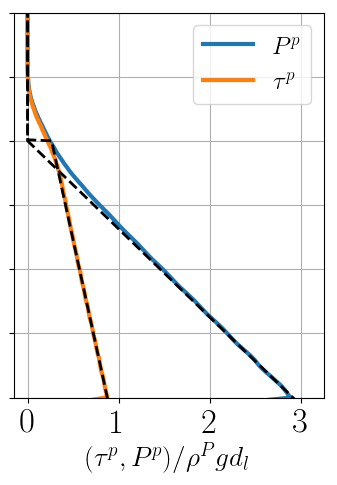}}
 \subfloat[]{\includegraphics[height=5.5cm]{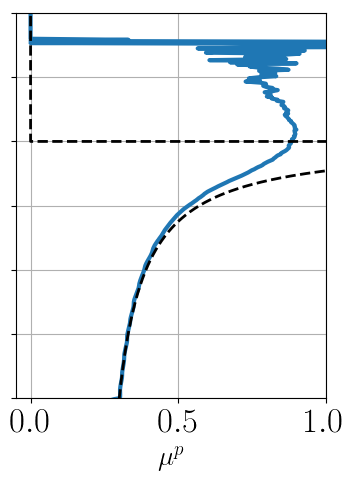}}
 \subfloat[]{\includegraphics[height=5.5cm]{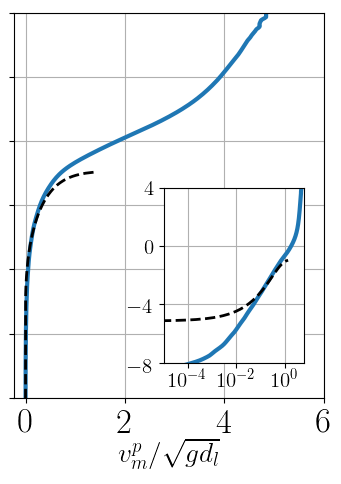}}
 \caption{Comparison between idealized (dotted lines) and DEM profiles (full lines) in the monodisperse configuration for $\theta \sim 0.45$. (a) Concentration profiles, (b) granular pressure and shear stress profiles, (c) friction coefficient profiles and (d) velocity profiles.}
 \label{fig:comp_stress}
\end{figure}

In practice, the concentration and velocity profiles, as well as the slip velocity, are difficult to obtain, and computing the additional transport due to the presence of small particles is not straightforward. In the following, a method to compute the two additional transport terms is proposed. The particles are assumed to be transported without dilatation of the bed. The concentration is therefore hypothesied constant and equal to $\phi_{max}=0.61$ in the bed with the top of the bed exactly at $z=0$ (see figure~\ref{fig:comp_stress}a). 
% \begin{equation}
% \phi_m(z) =
%  \left\{
%  \begin{array}{lr}
%     \phi_{max}=0.61, & z \leq 0 \\
%     0, & z >0
%  \end{array}
% \right.
% \label{eq:phi_id}
% \end{equation}
% 

To compute the $\Delta Q_1$ additional small particle transport term, the monodisperse velocity profile for $z \leq z_i$ needs to be estimated. It can be derived using the $\mu(I)$ rheology (equation~\ref{eq:muI}). The stress state (normal and shear stresses) of the granular bed needs also to be computed. Based on the two-phase volume-averaged equations for turbulent bedload transport \cite{jackson2000, chauchat2018} and for the idealized step concentration profile (figure~\ref{fig:comp_stress}a), the granular pressure and shear stress profiles can be expressed as (see appendix~\ref{sec:app1})
\begin{equation}
 P^p(z) = \left(\rho^p-\rho^f\right)g\cos(\alpha)\phi_{max}z,
 \label{eq:P2}
\end{equation}
\begin{equation}
 \tau^p = \tau_b + \left(\rho^f + (\rho^p-\rho^f)\phi_{max}\right)z,
 \label{eq:tau2}
\end{equation}
where $\tau_b =\rho^f g h_w \sin(\alpha)$ is the fluid bed shear stress. The friction coefficient can then be computed analytically as $\mu^p = \tau^p/P^p$ with these profiles. Inverting the $\mu(I)$ rheology (equation~\ref{eq:muI}), replacing the inertial number $I$ by its expression (equation~\ref{eq:Idef}) with the large particle diameter and integrating, a velocity profile is obtained 
\begin{equation}
v_m^p(z) = 
\left\{
 \begin{array}{lr}
  0, & \mu^p(z)<\mu_1,\\
  \displaystyle \int_{z_1}^z\left(\sqrt{\dfrac{P^p(\zeta)}{\rho^p}}\dfrac{I_0}{d_l}\dfrac{\mu^p(\zeta) - \mu_1}{\mu_2-\mu^p(\zeta)}\right)d\zeta, & \mu_1 \leq \mu^p(z)<\mu_2,
 \end{array}
 \right.
 \label{eq:vxP}
\end{equation}
where $\mu_1=0.35$, $\mu_2=0.97$ and $I_0=0.69$ are the set of parameters proposed by Maurin \textit{et al.} \cite{maurin2016} for bedload transport. The integral can be computed numerically with the analytical expression of the granular pressure and of the friction coefficient and without any data from the DEM simulations. 

To verify that this derivation is consistent with the DEM simulations, figure~\ref{fig:comp_stress} compares, for the monodisperse simulation at $\theta \sim 0.45$, (a) the idealized concentration, (b) the pressure and shear stress, (c) the friction coefficient and (d) the velocity profile with the DEM results. The idealized step concentration profile obviously does not reproduce the dilatation of the bed at the surface. As a result, the pressure and shear stresses correspond with the DEM results in most part of the bed but differ close to the surface. Similarly discrepancies near the bed surface appear for the friction coefficient and the velocity profiles. However, in the expression of $\Delta Q_1$, the velocity and concentration profiles are needed only for $z \leq z_i$, where the idealized concentration and stresses agree very well with the DEM ones. Concerning the velocity profile (figure~\ref{fig:comp_stress}d), the $\mu(I)$ rheology can not predict the quasi-static regime as already mentioned (see inset). The velocity profile is well predicted in the dense regime but the rheology fails to predict the velocity in the upper part of the bed for $\mu^p \geq \mu_2$, which corresponds to a more dilute flow regime. In order to use the predictive model, it is therefore necessary that $\mu^p(z_i) < \mu_2$, which is the case in all our simulations and should be the case in classical bedload transport configurations. Otherwise, it would mean that small particles are in the dilute flow regime and would be present at the bed surface, configuration which has already been discarded. With the velocity profile~\eqref{eq:vxP}, it is now possible to compute the first additional transport term $\Delta Q_1$ without any data from the DEM simulations.\\

To compute the second additional transport term $\Delta Q_2$, both the $\Delta v$ slip velocity and the $\int_{z_i}^{+\infty}\phi_m(z)dz$ term need to be estimated. The second term represents the amount of large particles slipping above the small particles. With the idealized concentration profile, it can be directly computed as
\begin{equation}
 \int_{z_i}^{+\infty}\phi_m(z)dz = \phi_{max}N_ld_l.
\end{equation}
Lastly, the slip velocity remains to be estimated. By definition, for $z\geq z_i$, $\Delta v = v_b^p(z) - v_m^p(z)$. It is therefore valid in $z=z_i$, where $v_b^p(z_i) = d_l/d_s v_m^p(z_i)$. The slip velocity is therefore finally given by
\begin{equation}
 \Delta v = \left(\dfrac{d_l}{d_s}-1\right)v_m^p(z_i),
 \label{eq:deltav}
\end{equation}
with $v_m^p(z_i)$ which can be computed from the velocity profile equation~\eqref{eq:vxP} derived previously. All additional transport terms can now be computed and the total additional transport can be expressed as
\begin{equation}
 \Delta Q = (\dfrac{d_l}{d_s}-1)\phi_{max}\left(\int_{0}^{z_i} v_m^p(z)dz + N_ld_lv_m^p(z_i)\right),
 \label{eq:minimod_deltaq}
\end{equation}
with $v_m^p(z)$ given by equation~\eqref{eq:vxP}. This additional transport term can be computed without any DEM data and uses only the $\mu(I)$ rheology.\\

\begin{figure}
 \subfloat[]{\includegraphics[height=6cm]{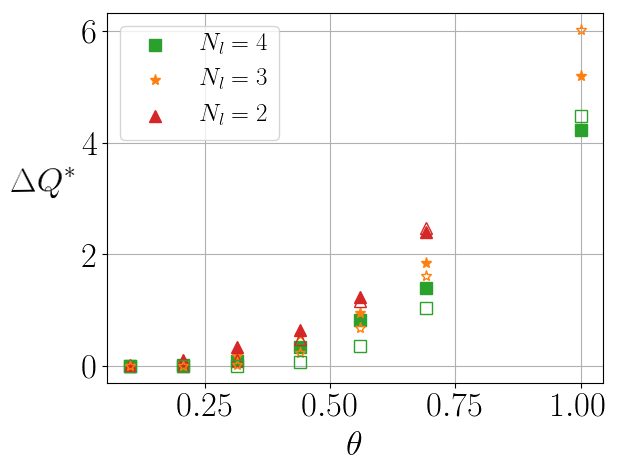}}
 \subfloat[]{\includegraphics[height=6cm]{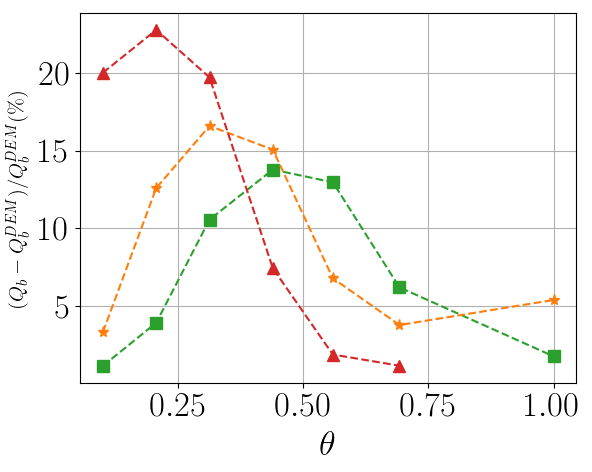}}
 \caption{(a) Dimensionless additional transport in the bidisperse case obtained with the DEM simulations (full symbols) and computed with equation~\eqref{eq:minimod_deltaq} (empty symbols), for different values of Shields number and $N_l$. Only cases for which the surface is composed of large particles only are presented for readability. (b) Error between the total bidisperse transport predicted by equation~\eqref{eq:minimod_deltaq} and the measured transport with the DEM simulations.}
 \label{fig:minimod_res}
\end{figure}

Equation~\eqref{eq:minimod_deltaq} is tested and compared with the additional transport rate directly obtained with the DEM simulations in figure~\ref{fig:minimod_res}. The model predicts well the additional transport with a maximum error around $20\%$, remaining smaller than $10\%$ in most cases. The error is generally smaller when $N_l$ is larger. For each configuration, there is a region where the error is maximum. The Shields number at which the maximum error is reached seems to depend on the large particle number of layers. These results are discussed and interpreted in the next section.

\section{Discussion and conclusion}\label{sec:discussion}

This study has shown that the additional transport evidenced in an inversely graded bidispersed bed is a granular process. In a granular flow, small particles, being more mobile than larger ones, play the role of a conveyor belt for the overlying large particles. Assuming that large and small particles are completely separated and are transported without dilatation of the bed, a model for the enhanced transport has been derived based on rheological arguments. The results have shown that our model contains the significant physical ingredients of the transport process and is able to predict acccurately the additional transport due to bidispersity in bedload transport. The developed model allows improving upon classical transport laws by taking into account not only the classical bed surface state, but the entire mobile granular bed structure.  \\

\begin{figure}
 \centering
 \includegraphics[width=\linewidth]{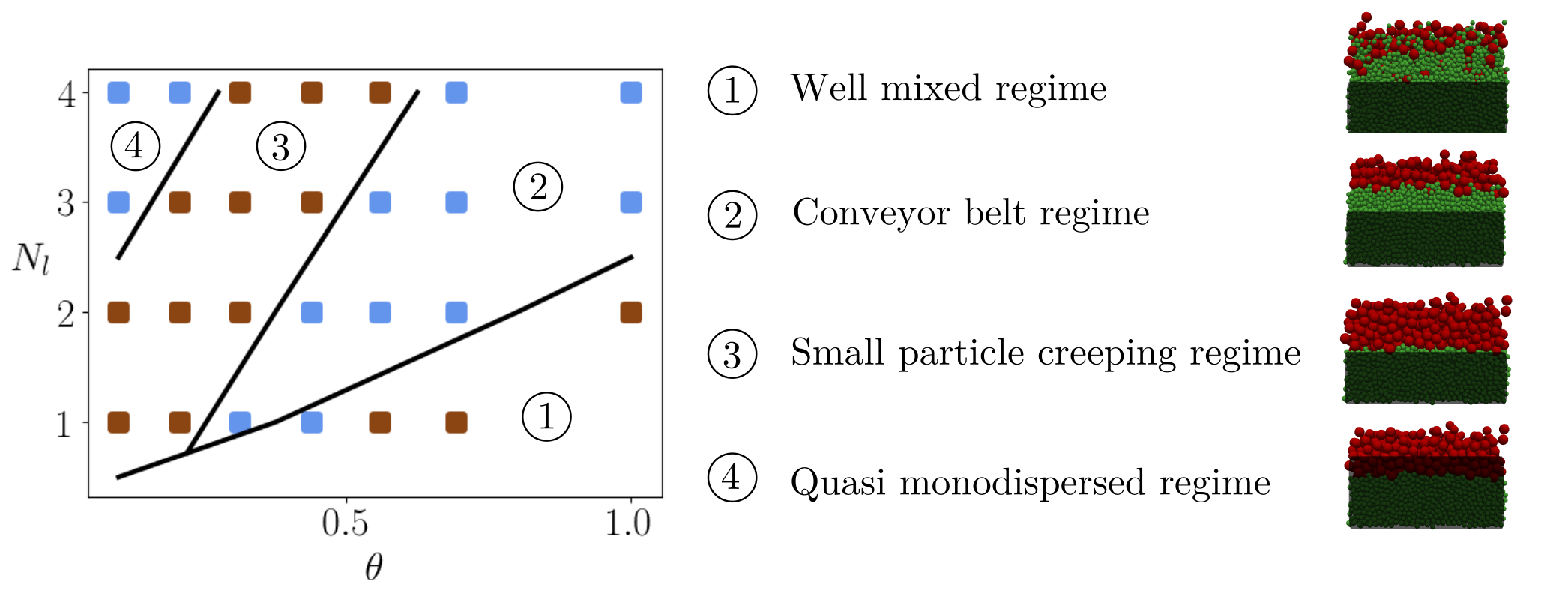}
 \caption{Mapping of the four different observed phenomenologies in the bidisperse transport process. Each regime is illustrated with a typical simulation picture where the creeping flow has been shaded in gray. Results are plotted in colored squares and split into two classes : blue (predicted transport error less than $10\%$) and brown (larger error).}
 \label{fig:phenom}
\end{figure}

This model can also be used as a tool to interpret the different transport mechanisms observed in this bidisperse granular flow configuration. The different regimes observed are summarized in figure~\ref{fig:phenom}. The map has been built from the regions of validity of the model, the blue squares showing regions where the error between the model prediction and the DEM is less than $10\%$ while the brown ones show regions where the error is higher. This criterion enables us to define four different regimes of granular flows, corresponding to different granular depth structure and flowing mechanisms. Regime 1 corresponds to cases where small and large particles are well mixed, with small particles present at the bed surface. In those cases, the additional transport is a combination of granular and fluid processes. Indeed, smaller particles at the surface are more easily entrained by the fluid flow and the mixture of small and large particles can affect the flowing properties of the granular mobile layer.  Regime 2 corresponds to the domain of validity of the proposed model, where all assumptions are verified. In this regime, the fluid-driven large particles entrain the small ones, which create a so-called conveyor belt effect, due to their higher mobility. The transition depth between small and large particles is here located in the dense granular flow region. When the transition is located deeper in the bed, near or inside the creeping flow region, the $\mu(I)$ rheology is no longer valid and the model predicts erroneously a zero velocity inside the small particle layer (see inset figure~\ref{fig:comp_stress}d). This third regime therefore leads to small ($<25\%$) but non negligible errors in the model predictions, due to the absence of slip velocity and additional transport. This indicates that the quasi-static part of the bed may play a non neglible role in the sediment transport process \cite{houssais2015, ferdowsi2017}. Regime 4 corresponds to cases where the transition depth is very deep in the bed and no additional transport due to the presence of small particles is observed in the DEM simulations or predicted by our model. The bidisperse nature of the bed can be neglected in this regime.

The model and the phenomenology map have been derived considering assumptions a priori valid for any granular flow on a pile. Therefore, this analysis should remain valid for other flow configurations of bidisperse mixtures with larger particles on top of smaller ones. In addition, the mechanisms described herein rely only on rheological arguments and one can expect the analysis to hold for any granular flow. Indeed, when submitted to the same stress, small particles are more mobile than larger particles and the effect observed for polydisperse granular collapses \cite{linares-guerrero2007, lai2017} or granular avalanches, for example, can be interpreted similarly. During the collapse, the small particles segregate and form a basal flowing layer, setting up a conveyor belt effect and increasing the runout distance of the collapse.\\
%According to the analysis presented herein, the small particle layer will act as a conveyor belt for the overlying large particles leading to an increase of the runout distance of the collapse.\\

\begin{figure}
\centering
 \subfloat[]{\includegraphics[width=.5\linewidth]{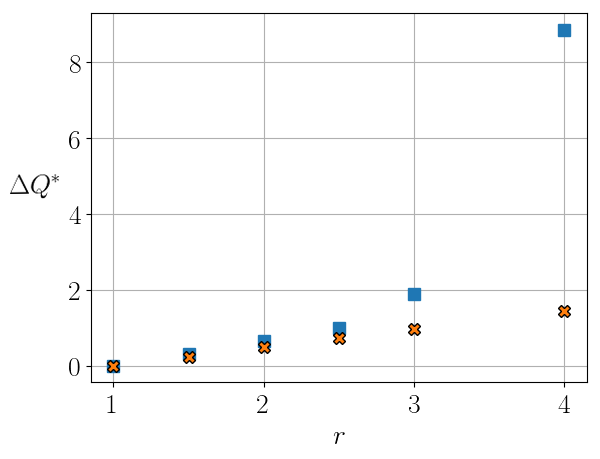}}
 \subfloat[]{\includegraphics[width=.5\linewidth]{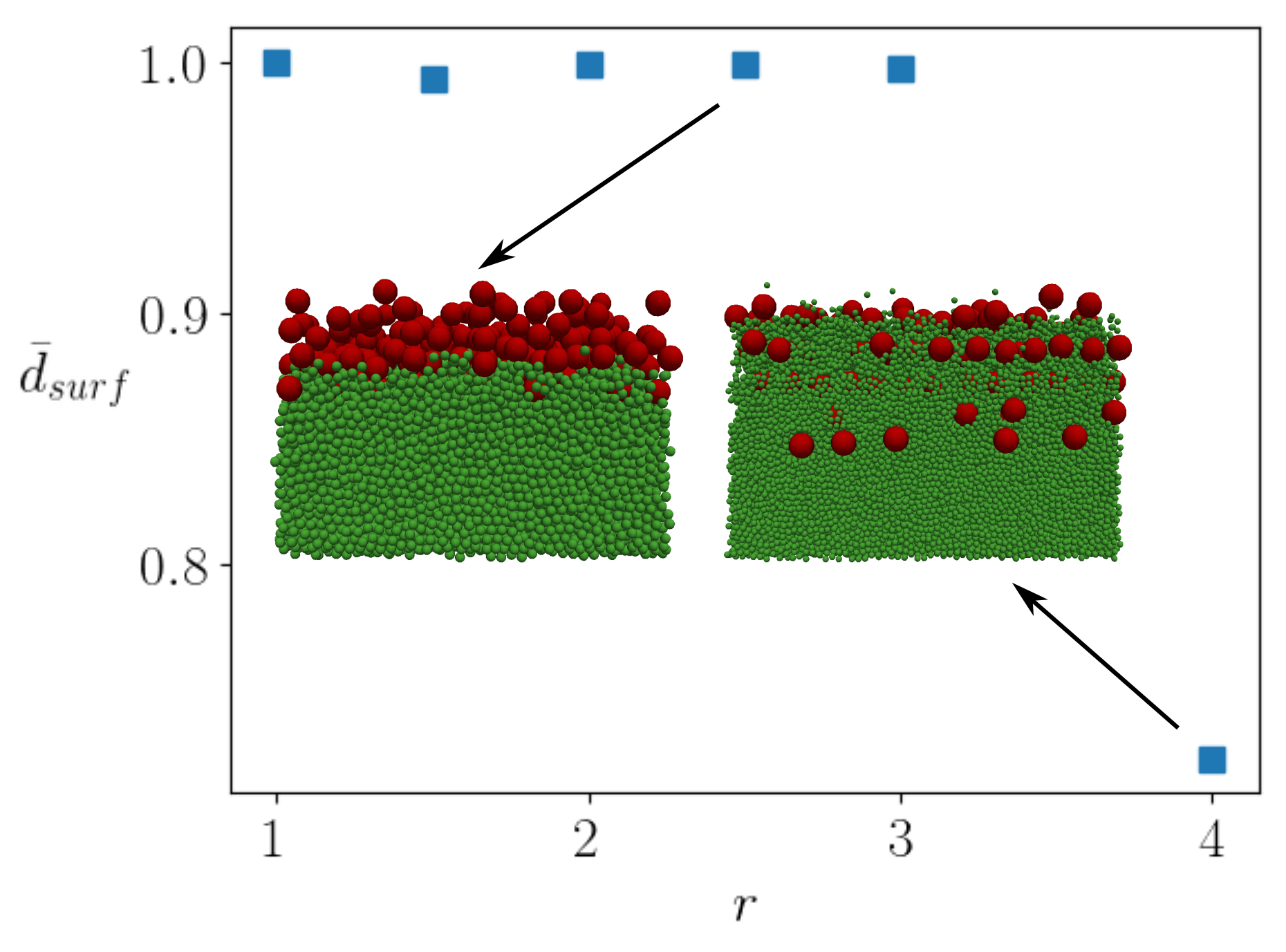}}
 \caption{(a) Additional transport rate predicted by equation~\eqref{eq:minimod_deltaq} (orange crosses) and computed from the DEM simulations (blue squares) for different size ratios at $\theta \sim 0.45$ and $N_l=2$. (b) Non-dimensional surface diameter in the DEM simulations as defined in equation~\eqref{eq:dsurf} as a function of the size ratio. Cases $r=2.5$ and $r=4$ are illustrated with a picture from the DEM simulation.}
 \label{fig:res_sizeratio}
\end{figure}

The results obtained in this study can be put into perspective by considering the dependency of the results on the size ratio. Varying the size ratio between $r = 1.5$ and $r = 4$ for a given configuration ($\theta = 0.45$, $N_l = 2$), one can evidence that the transport predicted by the model is valid up to $r = 2.5$ (see figure~\ref{fig:res_sizeratio}a). For a larger size ratio,  the increased transport observed in the DEM is much higher than predicted by the model. This effect seems to be related to a drastic change in the granular flow structure. Indeed, the mean surface particle diameter, representative of the mixing of small and large particles changes drastically between a size ratio of $r=3$ and $r=4$ (see figure~\ref{fig:res_sizeratio}b). This indicates that diffusion remixing increases significantly, and can be related to the onset of inverse segregation as observed in this range of size ratio by Thomas \cite{thomas2000}. This link between diffusion and inverse size segregation challenges our understanding of size segregation and deserves future work. \\

\section*{Acknowledgements}

This research was funded by the French Agence nationale de la recherche, project ANR-16-CE01-0005 SegSed 'size segregation in sediment transport'. The authors acknowledge the support of INRAE (formerly Irstea and Cemagref). INRAE, ETNA is member of Labex Osug@2020 (Investissements d’Avenir Grant Agreement ANR-10-LABX-0056) and Labex TEC21 (Investissements d’Avenir Grant Agreement ANR-11-LABX-0030).

We are grateful to M. Church for reviewing and English corrections.

\appendix 
\section{Derivation of the granular stress profiles}\label{sec:app1}

\begin{figure}[h!]
 \subfloat[]{\includegraphics[height=5cm]{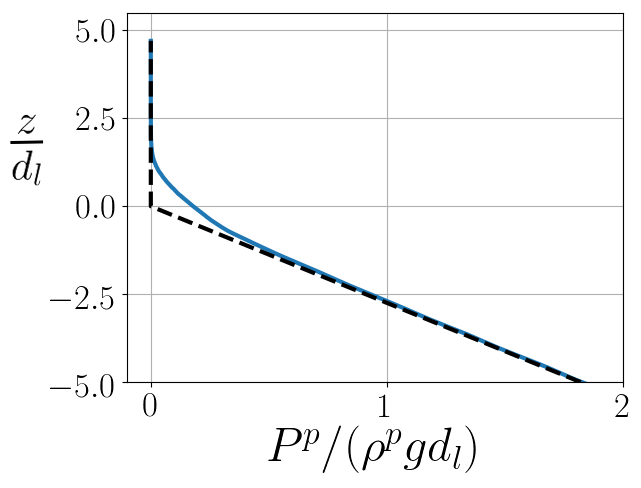}}
 \subfloat[]{\includegraphics[height=5cm]{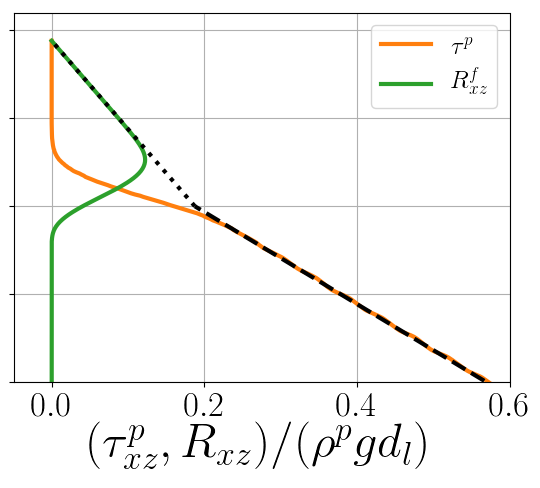}}
 \caption{Monodisperse case at $\theta\sim0.45$. (a) Granular pressure from DEM simulation (full line) and computed with analytical expression~\eqref{eq:pp} (dashed line). (b) Fluid and granular shear stress from DEM simulation (full line) and computed from analytical expression~\ref{eq:fluidStress} (dotted line) and expression~\eqref{eq:taup} (dashed line)}
 \label{fig:app}
\end{figure}

The two phase flow equations of bedload transport developed by \cite{revil-baudard2013} and \cite{chauchat2018} are considered. For a unidirectional flow and for steady state condition, they read
\begin{equation}
0 = \dfrac{\partial S_{xz}}{\partial z} + \dfrac{\partial R_{xz}}{\partial z} + \rho_f(1-\phi)g\sin(\alpha) -n\left<f_{f_x}^p\right>^s,
\label{eq:tauf_grad}
\end{equation}
\begin{equation}
0 = \dfrac{\partial \tau^p}{\partial z} + \rho^p\phi g \sin{\alpha} + n\left<f_{f_x}^p\right>^s,
\label{eq:taup_grad}
\end{equation}
\begin{equation}
0 = \dfrac{\partial P^f}{\partial z} + \rho^fg\cos{\alpha},
\label{eq:pf_grad}
\end{equation}
\begin{equation}
0 = \dfrac{\partial P^p}{\partial z} + (\rho^p-\rho^f)\phi g\cos{\alpha},
\label{eq:pp_grad}
\end{equation}
where $S_{xz}$ and $R_{xz}$ are the viscous and turbulent fluid shear stresses, $\tau^p$ is the granular shear 
stress, $n\left<f_{f_x}^p\right>^s$ represents the transfer of momentum from the 
fluid to the solid phase and $P^f$ and $P^p$ are the fluid and granular pressure.
\cite{maurin2015} showed that the viscous fluid shear stress $S_{xz}$ is negligible 
in the bedload configuration and it will therefore not be taken into account. Considering the following idealized concentration profile 
\begin{equation}
 \phi = 
 \left\{
 \begin{array}{cc}
  \phi_{max}=0.61, & \text{ if } z\leq 0,\\
  0, & \text{ if } z > 0,
 \end{array}
 \right.
 \label{eq:phi_id}
\end{equation}
and by integration of equation~\eqref{eq:pp_grad} between
an elevation $z$ and $0$ where $P^p(0)$ is assumed to vanish, the two phase flow model 
predicts hydrostatic pressure for the granular phase
\begin{equation}
P^p(z) = -(\rho^p-\rho^f)\phi_{max}g\cos(\alpha)z.
\label{eq:pp}
\end{equation}
Summing equation~\eqref{eq:tauf_grad} and~\eqref{eq:taup_grad},
a mixture momentum balance is obtained 
\begin{equation}
0 = \dfrac{\partial R_{xz}}{\partial z} + \dfrac{\partial \tau^p}{\partial z} + \left(\rho^f + (\rho^p-\rho^f)\phi\right)g\sin(\alpha).
\label{eq:momentum}
\end{equation}
In order to understand the partition between the fluid and granular stresses, equation~\eqref{eq:momentum}
is integrated between an elevation $z$ and the free water surface $h_w$ where both shear stresses are 
assumed to vanish, leading to
\begin{equation}
R_{xz}(z) + \tau^p(z) = \left(\rho^f(h_w-z) + (\rho^p-\rho^f)\int_{z}^{h_w}\phi(\xi)d\xi\right)g\sin(\alpha).
\label{eq:momentum2}
\end{equation}
In the pure fluid phase, where $\phi = 0$ and therefore $\tau^p(z)=0$, equation~\eqref{eq:momentum2} simplifies to
\begin{equation}
 R_{xz}(z) = \rho^fg\sin (\alpha) (h_w-z),
 \label{eq:fluidStress}
\end{equation}
the classical expression of the turbulent fluid shear stress in a free surface flow. 
In the granular bed the fluid shear stress rapidly decreases to zero and only
the granular shear stress holds the mixture shear stress. With the idealized concentration profile~\eqref{eq:phi_id}, 
equation~\eqref{eq:momentum2} simplifies to
\begin{equation}
\tau^p(z) = \left[\rho^f(h_w-z) - (\rho^p-\rho^f)\phi_{max}z\right]g\sin(\alpha),
\end{equation}
which can be rewritten as
\begin{equation}
\tau^p(z) = \rho^fg\sin(\alpha)h_w - \left[\rho^p\phi_{max} + (1-\phi_{max})\rho^f\right]gsin(\alpha)z.
\label{eq:taup}
\end{equation}
The expressions of the granular pressure, fluid shear stress and granular shear stress obtained for the idealized step concentration are compared with the DEM profiles in figure~\ref{fig:app}. They agree in most parts except in the transition from the compacted granular bed to the pure fluid phase that is not modeled by the idealized concentration profile. This step concentration profile corresponds to an idealized situation where the fluid shear stress is completely transmitted to the granular bed at the discontinuity ($z=0$). Focusing on the granular shear stress, the DEM and analytical profiles correspond almost perfectly as soon as $z \leq 0$ meaning that the fluid stress is indeed completely transmitted to the granular bed over a depth much smaller than a particle diameter.

\bibliography{biblio}

\end{document}